\documentclass[conference,10pt]{IEEEtran}
\usepackage{algorithm,multirow}
\usepackage[noend]{algorithmic}
\usepackage{graphicx}
\usepackage{comment}
\usepackage{cite}
\usepackage{caption}
\usepackage{multirow}
\usepackage{amssymb,amsthm,mathrsfs}
\usepackage{amsfonts,epsf,graphicx,times,amsmath,multirow,url,cite}
\usepackage{subcaption}
\usepackage{float}
\DeclareGraphicsExtensions{.eps,.pdf,.jpeg,.png}

\makeatletter
\def\ps@headings{%
\def\@oddhead{\mbox{}\scriptsize\rightmark \hfil \thepage}%
\def\@evenhead{\scriptsize\thepage \hfil \leftmark\mbox{}}%
\def\@oddfoot{}%
\def\@evenfoot{}}
\makeatother
\newcommand{\sol}{\texttt{EDOS}}
\newcommand{\ea}{\texttt{Edge Assistant}}
\newcommand{\oa}{\texttt{Offloading Agent}}
\newcommand{\eashort}{\texttt{EA}}
\newcommand{\oashort}{\texttt{OA}}

\begin{document}

\date{}


\title{\huge EDOS: Edge Assisted Offloading System for Mobile Devices}

\author{
\IEEEauthorblockN{Hank H. Harvey\IEEEauthorrefmark{1},
Ying Mao\IEEEauthorrefmark{1},
Yantian Hou\IEEEauthorrefmark{2} and
Bo Sheng\IEEEauthorrefmark{3} }
\IEEEauthorblockA{\IEEEauthorrefmark{1}Department of Computer Science,
The College of New Jersey, Email: \{harveyh1, maoy\}@tcnj.edu}
\IEEEauthorblockA{\IEEEauthorrefmark{2}Department of Computer Science,
Boise State University, Email: yantianhou@boisestate.edu}
\IEEEauthorblockA{\IEEEauthorrefmark{3}Department of Computer Science,
University of Massachusetts Boston, Email: shengbo@cs.umb.edu}
}

\maketitle
\begin{abstract}
Offloading resource-intensive jobs to the cloud and nearby users is a promising approach to enhance mobile devices.
This paper investigates a hybrid offloading system that takes both infrastructure-based networks and Ad-hoc networks into the scope.
Specifically, we propose EDOS, an edge assisted offloading system that consists of two major components, an Edge Assistant (EA) and Offload Agent (OA).
EA runs on the routers/towers to manage registered remote cloud servers and local service providers
and OA operates on the users' devices to discover the services in proximity. We present the system with a suite of protocols to collect the potential
service providers and algorithms to allocate tasks according to user-specified constraints.
To evaluate EDOS, we prototype it on commercial mobile devices and evaluate it with both experiments on a small-scale testbed and simulations.
The results show that EDOS is effective and efficient for offloading jobs.

\end{abstract}
\section{Introduction}

Nowadays, mobile devices and mobile apps have been seamlessly weaved into people's daily life.
Both hardware and software have evolved rapidly to fulfill the demands of the market.
Although the state-of-the-art mobile device hardware is capable of supporting a large set of various applications,
it is still limited compared to regular computers and servers, especially in terms of computation ability and network
bandwidth when considering the computational intensive tasks in imaging~\cite{yancy9175642, yancy2021form}, financial modeling~\cite{zhu2020high, jarrow2021low, zhu2021time, zhu2020adaptive}, on-device machine learning~\cite{zhu2021news, zhu2021clustering,li2021frequentnet}. Mobile users, however, desire some computation-intensive applications that may not be suitable for
mobile devices, e.g., popular cloud-side services like voice recognition, face recognition, and image/video rendering.
In addition, energy consumption is another critical hurdle for some applications to deploy on mobile devices.

Offloading is a well-accepted approach that helps overcome the resource limitation by
allowing a device with resource constraints to delegate its jobs or applications
to another powerful device for execution. The powerful device can be physically nearby or
remotely connected via the Internet. For mobile devices, the current infrastructure and
technology offer a wide range of choices as offloading targets including cloud-side servers,
other nearby mobile devices, emerging edge computing devices, and even IoT devices.
While the basic approach of offloading is straightforward, it is challenging to determine an
appropriate offloading plan that involves various types of devices.

\begin{figure}[ht]
\centering
\includegraphics[width=0.9\linewidth]{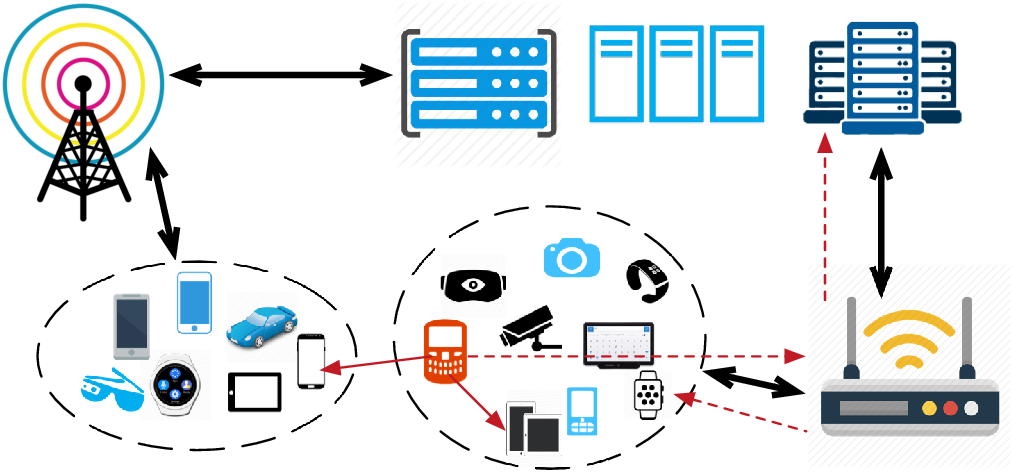}
\caption{\sol~in a heterogeneous network}
\label{fig:system-arch}
\end{figure}

In this paper, we develop a EDge assisted Offloading System~(\sol).
The system targets users'  mobile phones, pads and smart watches, as well as their
smart glasses or helmets with virtual or augmented reality, connected vehicles and various Internet of Things~(IoT) devices.
The main objective of the system is to select a set of devices to collaboratively and efficiently accomplish the job.
To construct a robust system, a user chooses potential nodes
that are willing to provide services from both nearby users and remote servers and then, offloads the jobs to selected nodes.
Fig.~\ref{fig:system-arch} shows an overview of the system with one user in red. This user utilizes two different networks to
discover nodes with services, the infrastructure-based network and the Ad-hoc network.
To discover remote service providers on the cloud, it accesses the network through a router (or cellular tower) and fetches
data from servers. In addition, the router can direct the user request to a local node that offers services (red dotted line).
At the same time, it can query the Ad-hoc network to discover nearby nodes that offer services(red solid line).

Our main contributions are as follows:
\begin{itemize}
 \item We propose \sol, an edge assisted offloading system that discovers
 services from both traditional infrastructure-based networks and Ad-hoc networks.
 \item We consider a dynamic job setting where a job can be split into a number of tasks which can be reassembled afterwards.
 The input and output size can be different.
 \item We mathematically formulate the problem and develop a suite of protocols along with algorithms to efficiently
 address it.
 \item We evaluate \sol~through popular applications on a small-scale test bed. The result shows a significant reduction
 in the average job completion time. Furthermore, we conduct simulations to evaluate \sol~in a large-scale environment.
\end{itemize}

\section{Related work}
With prevalence of computing infrastructures, mobile systems, such as smartphones, 
benefit from various emerging technologies. 
However, the limited onboard resources, such as battery life, network bandwidth, and storage capacity obstruct mobile devices from various applications.  
As a practical approach, offloading those resource-intensive jobs to the cloud or other users is gaining attention in the research communities.

Depending on the system design, offloading operations may be performed at different levels, such as methods~\cite{corral2014method}, tasks~\cite{imai2011light}, 
applications~\cite{xian2007adaptive}, virtual machine~\cite{hutchins2012offloading} and code~\cite{kosta2012thinkair}. 
A prerequisite to an efficient offloading system is to decide which components to offload. 
Such decisions are based on the profiling data about application execution and system contexts, 
such as the CPU usage, energy consumption, and network latency~\cite{barbera2013offload}. For example, MAUI~\cite{cuervo2010maui}
provides a system framework that enables energy-aware offloading of mobile
code to the infrastructure. However, MAUI system relies on developers’ efforts to annotate the methods that should be offloaded. 
On the other hand, CloneCloud~\cite{chun2011clonecloud} boosts unmodified mobile applications by seamlessly offloading part of their execution from the
mobile device onto device clones operating in a computational cloud. It determines these pieces with an offline static analysis of different running
conditions of the process' binary on both a target smartphone and the cloud.  
By deploying a Software Defined Network framework in the core mobile network, 
SMORE~\cite{cho2014smore} architecture allows offloading selected traffic to an in-mobile-core-network cloud platform without requiring protocol changes. 
Saving energy to extend the battery life is an important objective of the offloading
systems. Karthik et.al~\cite{kumar2010cloud} proposes an analytical model for comparing energy usage in the cloud and the mobile device.

Besides determining which components to offload, another aspect is where should the offloadable tasks go.
The MobiScud~\cite{wang2015mobiscud} system offloads these tasks to a personal cloud. In addition, 
it takes the mobility into consideration and ensures a low latency between mobile devices and
cloud platforms is maintained as users move around. The authors in~\cite{cheng2014vehicular, wan2014vcmia, li2014coding} investigate the offloading system by using
the vehicle network to enable the data transmission between vehicles and
infrastructures. 
Opportunistic networks have also been studied for mobile offloading 
systems~\cite{han2012mobile,wang2014mobile, wang2014toss, mao2014pasa, mao2016mobile, mao2015building}.

A recent trend in the field is to enable Mobile Edge Computing (MEC). Several approaches have been trying to push the jobs to the edge. 
Chen et al.~\cite{chen2016efficient} proposed a distributed computational offloading model that 
uses game theoretical approach to achieve the Nash equilibrium of the multi-user computation offloading game.
Moreover, a dynamic computation offloading policy for MEC systems with mobile
devices powered by renewable energy is presented in~\cite{mao2016dynamic}. 
However, the unstable wireless connection between the edge and users results in a substantial delay.

Unlike the previous work, in this paper, we focus on developing an offloading system that considers a heterogeneous network. In our setting, the users hold various types of devices, regarding hardware, software and network association (e.g. cellular, WiFi). Additionally, the user can utilize both edge assistance to discover the potential service providers
and Ad-hoc networks to find nearby service nodes.
\section{Framework of \sol}

In this section, we present the details of \sol~system.
It mainly includes two components, \ea~(\eashort) and~\oa~(\oashort), where~\eashort~operates on routers or towers and~\oashort~
runs on users. \ea~and \oa~ are designed to gather the information, analyze the data and process the requests.

\subsection{Edge Assistant}
The \eashort~is a lightweight middleware that is running on cellular towers and routers. 
Due to unpredictable delays from users to various remote servers, deploying \eashort~on the edge of the wired network can reduce the workload of discovering the services on the user side. Additionally, some of the clients associated with this tower or router may also act as service providers. 
Therefore, the primary responsibilities of \eashort~include the management of 
registered remote service providers and clients that are connected to itself.
Fig.~\ref{fig:casho}(upper level) illustrates the major components in the architecture of \eashort, service manager and client manager.

\begin{itemize}
 \item {\bf Service Manager} is in charge of the coordination with registered service providers. First, for each provider, it collects the types of services it offers, the currently available resources as well as the delays to the remote servers. Due to rapid changes, this information needs to be updated timely. Then, it creates a virtual platform which includes the metadata of different providers.
 Whenever an offloading request arrives at service manager, it uses this virtual platform to estimate the cost for a user
under each particular remote server. 
 \item {\bf Client Manager} is a background service that constantly interacts with its host (towers and routers). 
 First, it fetches the current active users that are associated with this host.
 A user can identify itself as a service node which means it is willing to share the resources with nearby users. The client manager maintains a table for each of the service nodes. 
 This table contains the state information of service nodes, e.g., battery life percentage, network bandwidth, computation resources and delays. When the user leaves the network, e.g. moving out of the towers or routers, the table will be updated accordingly.
 Client manager uses this table to predict the cost to use services provided different nodes. 
\end{itemize}

\subsection{Offloading Agent}
\begin{figure}[ht]
\centering
\includegraphics[width=0.9\linewidth]{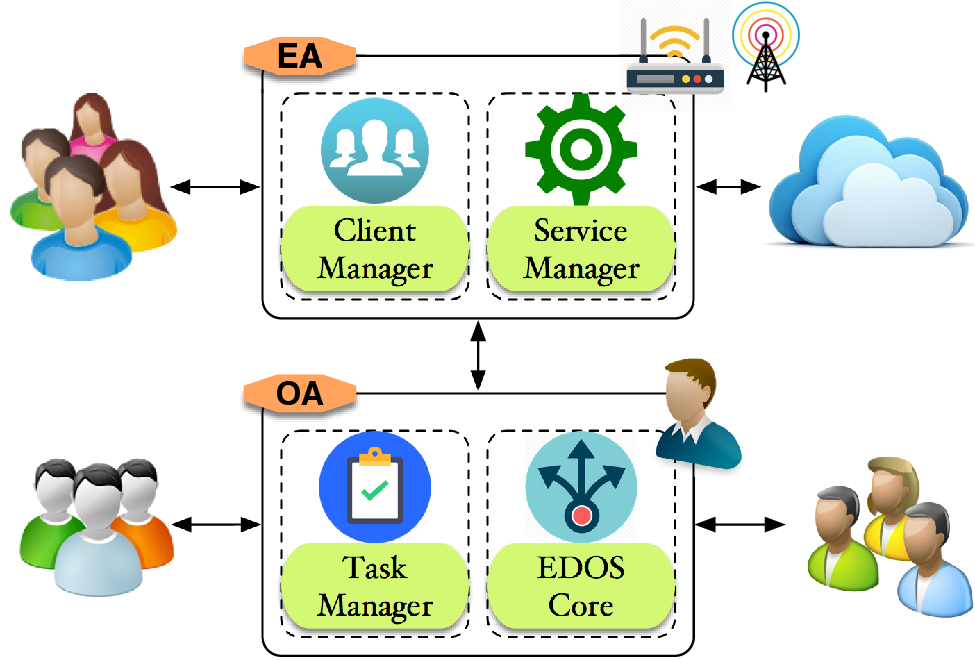}
\caption{Major components in \sol}
\label{fig:casho}
\end{figure}

The \oashort~is developed to operate on the users' devices to perform the essential functionalities, such as job analysis, service discovery and
task allocation, in \sol. Fig.~\ref{fig:casho}(lower level) presents \oashort~that consists of two principal components, job manager and \sol~core.
 
\begin{itemize}
 \item {\bf Job Manager} handles users' requests from the applications. If a request contains an offloadable job,
 this job can be further split into a number of tasks. Such a task is a minimum unit that can be processed by other nodes.
 The job manager maintains a table of the offloadable jobs and their corresponding tasks. 
 Each task contains an estimation of required resources and a budget that shows how much the user is willing to pay, 
 in terms of computation, bandwidth and/or money. 
\item {\bf \sol~Core} is a decision maker whose main responsibilities are discovering the service nodes and determining 
which service nodes should be selected for tasks. For service discovering, if the user is connected to an infrastructure-based network, 
it queries its \eashort~to fetch available service providers on the cloud and the nearby service nodes that associate with the same \eashort. 
Furthermore, it uses the Ad-hoc network to discover the nearby users who are willing to offer services but not within the same \eashort. 
After the discovery, the user generates two tables, one is node candidates with estimated time delays that include 
both computational and transmission delays, the other one is node candidates with their cost to complete the tasks.
 Based on these tables, \sol~core makes the decision on which candidates would be selected to perform the task. 
 The objective is to minimize time overhead, in the meanwhile complete task within the budget. 
\end{itemize}

\section{\sol~Service Discovery}
Previously, we discussed that the first step of any user in \sol~is to discover the service providers
for offloading. In this section, we present the service discovery protocol in our solution, \sol,~
which mainly consists of two separate parts: discovery with \ea~and discovery with Ad-hoc Networks.

\subsection{Discover Service Nodes with \eashort}
\begin{figure}[ht]
\centering
\includegraphics[width=0.65\linewidth]{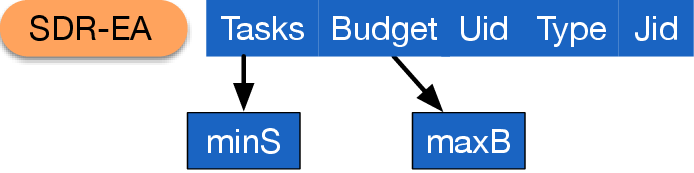}
\caption{Message format of SDR-EA}
\label{fig:sdr}
\end{figure}

\ea, running on the routers and towers, gathers the information of the clients that are associated with it and the remote servers that are registered with it.
The information which stores in a set $\{R\}$ includes service type, cost per unit and available resources, such as computation and bandwidth.
In general, \eashort~maintains $u_i\in U$ and $m_i \in M$ where $u_i$ is a
user with id $i$ and $m_i$ is the cloud service provider with id $i$.
Whenever the user has an offloadable job, it constructs a Service Discovery
Request for \eashort({\bf SDR-EA}). Fig.~\ref{fig:sdr} shows the format of a {\bf SDR-EA} message that contains its own user id ($Uid$), requested service type ($Type$),
job id ($Jid$), $Tasks$ and $Budget$. The task field stores the minimum size among split tasks ($minS$). The budget field includes the maximum budget
in the tasks ($maxB$).

Algorithm.~\ref{alg:ea} shows how the {\bf SDR-EA} is handled by \eashort. First of all, \eashort~maintains a $(Uid, Jid)$
pair and stores it into
a set, $\{A\}$, which can be used to identify the active offloading jobs and manage the total workload through the cardinality of $\{A\}$(lines 1-8).
For a remote server to be selected as a candidate for a particular job, it needs to satisfy the following conditions:
1) The service type, such as network-intensive and computation-intensive,
must match the job's $Type$. 2) The cost per unit can not exceed the maximum budget for a task;
otherwise, it cannot take any of the tasks in this job. Upon finding a satisfied server, the server's information set $R_c$ will be stored in $\{CAND\}$ (lines 9-11).
Following the same procedure, we enumerate the nearby service providers that connect to \eashort. In addition to the requirements for wired cloud servers, this provider, as a wireless node, should offer a larger bandwidth than the minimum task size.
Otherwise, it can not take a single task. After checking requirements, the \eashort~adds the candidate's {$\{R_u\}$} to $\{CAND\}$ (lines 12-14). Finally,
\eashort~updates the $\{A\}$ and returns the $\{CAND\}$ set to the requester (lines 14-16).

\begin{algorithm}[ht]
\begin{algorithmic}[1]
\STATE Maintains $\{A\}$ that stores activated offloading user ids and job ids
\STATE Candidates set $\{CAND\} = \emptyset$
\STATE {\bf function Receive(DSR-EA):}
\STATE Read $Uid$, $Type$, $Jid$, $Tasks$ and $Cost$ from {\bf SDR-EA}
    \IF{$(Uid, Jid)\notin \{A\}$ }
      \STATE Add $(Uid, Jid)$ to $\{A\}$

    \ELSE
      \STATE {\bf Return} Still Active
    \ENDIF

    \FOR {$m_i \in M$}
        \IF{$m_i.type == Type$ ~~\&\&~~ \\ $m_i.cost\times Tasks.minS < Budget.maxB$
        }
              \STATE Add $R_{m_i}$ into $\{CAND\}$
        \ENDIF
    \ENDFOR

    \FOR {$u_i \in U$}
        \IF{$u_i.type == Type$ ~~\&\&~~\\ $u_i.cost \times Tasks.minS < Budget.maxB$ ~~\&\&~~ \\
        $u_i.bandwidth > 2\times Tasks.minS$}
             \STATE Add $R_{u_i}$ into $\{CAND\}$
        \ENDIF
    \ENDFOR
    \STATE Remove $(Uid, Jid)$ from $\{A\}$
    \STATE {\bf Return $\{CAND\}$}
\end{algorithmic}
\caption{Process Service Discovery Request on \eashort}
\label{alg:ea}
\end{algorithm}

\subsection{Discover Service Nodes with Ad-hoc Networks}
\begin{figure}[ht]
\centering
\includegraphics[width=0.70\linewidth]{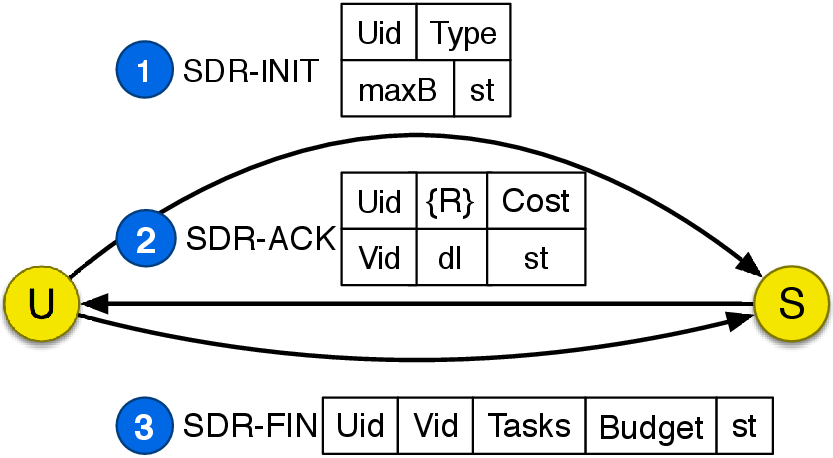}
\caption{Three-way handshake protocol}
\label{fig:handshake}
\end{figure}

In addition to infrastructure-based discovering, \sol~supports finding the nearby service nodes through Ad-hoc networks.
The nearby users can use the onboard Bluetooth or WiFi Direct modules to construct an Ad-hoc network.
Due to missing a centralized controller, like \eashort, when requesting the services, it is unlikely the user
has an updated list of nearby service nodes on hand. Therefore, we design a three-way handshake protocol to request services.
In the protocol, a user broadcasts out the {\bf SDR-INIT} message that contains its id ($Uid$), requested services type ($Type$), the maximum
budget for a split task ($maxB$) and the timestamp ($st$). Once receiving the message, targeted service nodes reply to it with a
{\bf SDR-ACK} message which consists of the requester's id ($Uid$), its own id ($Vid$) and resources set $\{R\}$,
such as id, computation and etc, cost per unit ($Cost$), the
delay between them ($dl$) and current timestamp ($st$). If a service node has been selected, the user sends out a {\bf SDR-FIN} message that includes
its id ($Uid$), target node id ($Vid$),
the split tasks ($Tasks$), the budget for each task ($Budget$), and current timestamp ($st$).
The structures of these three messages are illustrated in Fig.~\ref{fig:handshake}.

In the system, a user can act as a requester and a service node simultaneously.
Each node $v_i$ maintains a set $\{L\}$, which stores a list of nearby nodes, their service types, and delays.
A $\{CAND\}$ will be created if $v_i$ is a requester.
When overhearing the three-way handshake messages, every node applies the following Algorithm~\ref{alg:handshake}.
First, it initializes the parameters and reads the message to determine the type (lines 1-4). If it is a {\bf SDR-INIT},
it checks the requested service type with its types. If it finds the match, it then checks the maximum per task budget. The cost should be less than
this budget; otherwise, it can not take any of the tasks. If $v_i$ identifies itself as a target of this {\bf SDR-INIT}, it constructs
the {\bf SDR-ACK} message that includes $Uid$, $Vid$, $\{R\}$, $\{Cost\}$, $dl$ and $st$ (lines 5-7).
Upon receiving a {\bf SDR-ACK}, $v_i$ first checks whether this message is targeted on itself. If it is, $v_i$ calculates
the roundtrip delay of {\bf SDR-INIT} and {\bf SDR-ACK} and then, adds the information into both $\{CAND\}$ and $\{L\}$ (lines 8-11).
If it is not, $v_i$ computes the one-way delay from $Vid$ to itself and adds it into $\{L\}$ (lines 12-14).
When the arrived message is {\bf SDR-FIN}, $v_i$ checks whether it is the destination of this message. If it is, $v_i$ extracts
the $Tasks$ list and starts processing them. Otherwise, $v_i$ calculates the one-way delay between the sender and itself and stores it
into $\{L\}$ (lines 15-20).

\begin{algorithm}[ht]
\begin{algorithmic}[1]
\STATE At node $v_i$ with nearby nodes stored in $\{L\}$
\STATE $\{CAND\} = \emptyset$
\STATE {\bf function Receive(Msg):}

\STATE Read Msg

\IF{Msg is a {\bf SDR-INIT} message}
    \IF {$Type == v_i.Type$ and $maxB > v_i.cost$}
      \STATE Return {\bf SDR-ACK}
    \ENDIF

\ELSIF{Msg is a {\bf SDR-ACK} message}
    \IF{$Uid == v_i.id$}
    \STATE $Delay = timestamp - st + dl$
    \STATE Add ($\{R\}$, $Cost$, $Delay$) to $\{CAND\}$ and $\{L\}$

    \ELSE
    \STATE $Delay = timestamp - st$
    \STATE Add ($\{R\}$, $Cost$, $Delay$) to $\{L\}$
    \ENDIF

\ELSIF{Msg is a {\bf SDR-FIN} message}
  \IF {$Vid == v_i.id$}
  \STATE Extract $Tasks$ and start executing them
  \ELSE
    \STATE $Delay = timestamp - st$
    \STATE Add ($Uid$, $Delay$) to $\{L\}$
  \ENDIF

\ENDIF
\end{algorithmic}
\caption{Process Handshake Messages on \oashort}
\label{alg:handshake}
\end{algorithm}

\section{Task Allocation}
\label{sec:problem}


Given the candidate node sets $\{CAND\}$, along with their parameters $Delay$ and $Cost$,  we could present our task model. Specifically, we use $ v_j $ to denote each candidate node, and $ D_j^t , D_j^c$ as the delay, and $ C_{j}^t, C_{j}^c $ as the costs.

\subsection{Problem Formulation}
We first present the network model, task model and then formulate the task allocation problem. The major notations are listed in Table. \ref{table1}.

\begin{table}[htb]
	\centering
	\caption{Notations}
	\scalebox{0.95}{

		\begin{tabular}{ | c | l | }
			\hline

			$ S_l $     &   task $ l $'s size  \\ \hline
			$ D_j^c/C_j^c	 $     &  node $ v_j $'s computational delay/cost per unit data  \\ \hline
			$ D_j^t/C_j^t	 $     &  transmission delay/cost per unit data towards node $ v_j $ \\ \hline
			$C_{j,th} $      & cost upper limit on node $ v_j $ \\ \hline
			$ Q_{j,n} $	      &  node $ v_j $'s availability at time slot $ n $   \\ \hline
			$ d_l $	      & delay variable, caused by task $ l $  \\ \hline
			$ e_{l,j} $	      & task assignment variable of task $ l $ on node $ v_j $    \\ \hline
			$ a_{l,n} $	      & task $ l $'s starting  time slot \\ \hline
            $ T_n $	      & starting time of slot $ n $ \\ \hline
            $ T_s $	      & constant system time overhead \\ \hline

		\end{tabular}
	}

	\label{table1}
	\vspace{-5pt}
\end{table}



\subsubsection{Task Modeling}

Without loss of generality, we assume node $ v_0 $  generates the tasks and offloads them to other devices in the network. Let $ K $ be a job generated by $v_0$.
In our settings, each job $K$ can be split into $ L $ small tasks, $k_1, k_2,... k_L $. Each  task $k_l$ can be offloaded
by \oashort~to any of the $ J $ devices $v_j \in \{v_0, v_2, ... v_J \} $, including \oashort~itself, $ v_0 $. Note that the candidate devices consist of mobile nodes and cloud servers. The tasks are sequentially disseminated but could be processed by multiple nodes in parallel. Time is slotted into $ N $ pieces with fixed length $ \Delta T $, i.e., $ T_n-T_{n-1} =\Delta T, \forall 1\leqslant n \leqslant N $.

The \oashort~works based upon the input parameters that are generated from the raw data collected by EA and locally.
Such parameters include:  $ \mathbf{S} = (S_{l}) \in \mathbb{Z}^{L}$ denoting all tasks' sizes; $ \mathbf{D}^c = (D_{j}^c) \in \mathbb{R}^{J}$ and $ \mathbf{C}^c = (C_{j}^c) \in \mathbb{R}^{J}$ as the computational delay and cost rates on node $ v_j $; $ \mathbf{D}^t = (D_{j}^t) \in \mathbb{R}^{J}$ and $ \mathbf{C}^t = (C_{j}^t) \in \mathbb{R}^{J}$ being the transmission delay and cost rates towards $ v_j $. Here we assume the transmission time delay is small compared with the slot length $ \Delta T $. The cost upper limit is denoted as $ \mathbf{C}_{th} = (C_{j,th}) \in \mathbb{R}^{J}  $; The dynamic node availability status is $ \mathbf{Q} = (Q_{j,n}) \in \{0,1\}^{J \times N} $.








To guarantee every task is processed and all tasks are sequentially disseminated, we must have:
\begin{equation}
\sum_{n} a_{l,n}=1 ,  \forall l, \quad \sum_{l} a_{l,n} \leqslant 1, \forall n \label{constraint2}
\end{equation}


The time overhead $ d_l $ caused by task $ l $ can be denoted as:
\begin{equation}
d_l=\sum_{j} e_{l,j}(D_{j}^t \cdot S_l + D_{j}^c \cdot S_l)  , \quad \forall l \label{constraint3}
\end{equation}

 In constraint $(\ref{constraint3}) $,  the binary variable $ e_{l,j} $ denotes whether node $ v_j $ is chosen to process task $ l $. Since only one node is used to process each task $ l $, we have the constraint:

\begin{equation}
\sum_{j} e_{l,j}=1, \quad \forall l \label{constraint4}
\end{equation}

In addition, each node's overall  task assignment should not be beyond its computing capacity. Using $ C_{j}^c $ to denote the computational cost rate incurred by processing tasks on node $ v_j $, we have the node capacity constraint for each node $ v_j\mid_{j\neq 0} $:

\begin{equation}
\sum_{l}e_{l,j} \cdot C_{j}^c \cdot S_l < C_{j,th} \forall j \neq 0 \label{constraint5}
\end{equation}

Here the parameter $ C_{j,th} $ denotes the capacity upper limit at node $ j $.
For node $ v_0 $, the non-negligible transmission cost should be taken into account. Using $  C_j^t $ to denote the average transmission cost rate toward node $ v_j $, we should have:

\begin{equation}
\sum_{l} (e_{l,0} \cdot C_{0}^c \cdot S_l + \sum_{j \neq 0 } e_{l,j} \cdot C_j^t \cdot S_l) < C_{0,th}  \label{constraint6}
\end{equation}

Using binary parameter $ Q_{j,n} $  to denote node $ v_j $'s dynamic availability at any time slot $ n $, we use the following constraint to guarantee each task $ l $ is only assigned to the node $ v_j $ that is available at any time slot:

\begin{equation}
 Q_{j,n} \geqslant \sum_{l}   e_{l,j} \cdot a_{l,n}   \quad \forall n,j     \label{constraint7}
\end{equation}

\subsubsection{ Task Dissemination Problem Formulation}

Given all the input parameters, we can now formulate our problem. Our task dissemination problem is to find a device allocation scheme $ (e_{l,j}) \in \{0,1\}^{L \times J}   $ and a scheduling scheme $ (a_{l,n}) \in \{0,1\}^{L \times N}   $ that jointly minimize the overall time delay at node $ v_0 $ while satisfying all constraints. The mathematical formulation is shown as follows:

\begin{equation*}
		\begin{aligned}
			 \text{At node $ v_0 $ } \notag\\
			 {\text{minimize:}} \quad &   T_s+  \max_{l} \{ \sum_{n} a_{l,n} (T_n+d_l)  \}  \\
			 \text{s.t.} \quad 
			 &\text{scheduling definiteness (\ref{constraint2}) } \\
			&  \text{delay definition  (\ref{constraint3}) } \\
			&  \text{allocation definiteness  (\ref{constraint4}) } \\
			&  \text{node capacity  (\ref{constraint5}, \ref{constraint6}) } \\
			&  \text{node dynamic availability  (\ref{constraint7}) } \\
		\end{aligned} \label{formulate}
\end{equation*}

In the objective function, $T_s$ is the constant dividing time  overhead for job $K$.    $ T_n $ is the total elapsed time before slot $ n $. Our problem is a mixed integer nonlinear programing (MINLP) problem, which is NP-hard in general.


\subsection{Task Allocation Algorithm Design}
In this subsection, we present an efficient algorithm to solve the task allocation problem.
Our objective is to utilize the information in the $\{CAND\}$ set to select service nodes for all tasks.
The total cost should be less than the user's preset budget and the job should be completed as soon as possible.
Recall that we split a job into multiple tasks. These tasks may be correlated with each other, i.e. Google Street View application
discussed in section~\ref{eval}. We define a correlated priority function, $P(k_i,k_j)$, where $k_i, k_j \in K$.
$P(k_i,k_j) = 1$ means tasks $k_i$ and $k_j$ have the priority to be allocated to the same service provider, otherwise, $P(k_i,k_j) = 0$.

Running on \oashort, Algorithm~\ref{alg:oa} assigns the tasks to candidate service nodes.
First, the \oashort~sorts the candidate set by the product of the cost and delay. Then it initializes the parameters $id, i, m$ and the ordered task set
$\{OT\}$ (line 1-2). After initialization, starting from $k_i$, it enumerates the elements in task set $\{K\}$ to find the correlated $k_m$. Then
$k_i$ and $k_m$ are assigned with continuous $id$, loaded to the ordered task set $\{OT\}$ and removed from $K$. This process is repeated until all tasks are sorted.
 When $|K|=0$, the set $\{OT\}$ contains all the ordered tasks (line 3-12).
For each service provider, $v_i$, in sorted candidate set, we feed it with tasks until the budget limit is reached. Since $v_i$ has a budget of cost (prevent resources draining out on one user), the algorithm needs to check if there is still room for the task
before allocating it (line 13-18). We remove $ v_j $ from the $\{CAND\}$ set whenever it is out of space for additional tasks (line 19-21).
After the task allocation, if $|OT|>0$, meaning the algorithm fails to find an appropriate service provider to meet the budget, then all the remaining tasks will be executed locally (line 22-23).

\begin{algorithm}[ht]
\begin{algorithmic}[1]
\STATE Sort candidates by $C_j^c\times(D_j^c + D_j^t)$ in an increasing order
\STATE Initialize $id, i, m$, $\{OT\}=\emptyset$ (Ordered Tasks set)
      \WHILE {$|K|>0$}
      \FOR{$k_i \in K$}
      \STATE $k_i.id = id$
      \FOR {$k_m\in K$}
      \IF{$P(k_i, k_m) == 1$}
	 \STATE $k_m.id = ++id$
	 \STATE $i=m$
	 \STATE  Add $k_i$, $k_m$ into $OT$
	  \STATE Remove $k_i$, $k_m$ from $K$
	  \STATE {\bf Break}
      \ENDIF
      \ENDFOR
      \ENDFOR
      \ENDWHILE

  \FOR{$v_j \in \{CAND\}$}
      \FOR{$k_i\in OT$}
	  \IF{$k_i.budget < k_i.size\times C_j^c$ and $C_{j,B} > 0$}
	  \STATE $C_{j,B}= C_{j,B} - k_i.size\times C_j^c$
	  \STATE $e_{i,j} = 1$
	  \STATE Remove $k_i$ from $OT$
	 \ELSE
	  \STATE Remove $v_j$ from $\{CAND\}$
	  \STATE Break
	  \ENDIF
       \ENDFOR
  \ENDFOR
  \IF {$|OT| > 0$}
  \STATE Execute the unassigned tasks locally
  \ENDIF
\end{algorithmic}
\caption{Task Allocation in \sol~system}
\label{alg:oa}
\end{algorithm}

\section{Implementation and Evaluation}
\label{eval}
In this section, we will first introduce the workloads which we used to test our \sol~system,
then discuss the implementation of \sol~ and finally present the performance
evaluation results from both experiments on a small-scale testbed and simulations.

\subsection{Understanding the Workloads}
In our problem settings, each offloadable job generated by the user can be split into several tasks for the further process. This is a commonly applied setting in many fields, such as in virtual reality, which usually involves panoramic photos from a 360-degree camera.
Google street view is another representative use case of~\sol.
It provides panoramic views from positions along many streets for more than 70 countries and 6000 cities.
Google produces the street views in three steps: firstly, the street-view vehicle that is equipped with multiple cameras
drives around and photographs the locations; secondly,
it combines signals from sensors on the vehicle that measure GPS, speed, and direction
to match each image to its geographic location on the map; finally, it applies image processing algorithms to
‘stitch’ the small photos together into a single 360-degree image where those small photos taken by adjacent cameras are slightly overlapping each other. Consequently, loading a street view is an offloadable job
and those small photos are tasks that split from such a job.


Regarding the image quality, the street view offers 5 levels, and each level corresponds to a number of small images. From level 1 to 5, the number of
small images is 2, 8, 28, 91, and 338, respectively.
Each level has a default resolution, which are 832$\times$416, 1664$\times$832, 3328$\times$1664, 6656$\times$3328 and 13312$\times$6656, respectively.
To utilize the Google street view, the user needs to download the small pictures and stitch them into a panoramic photo.
When stitching, the user can specify an appropriate resolution that is suitable for this device.


\begin{figure}[h]
   \centering
      \begin{subfigure}[t]{0.47\linewidth}
\centering
      \includegraphics[width=\linewidth]{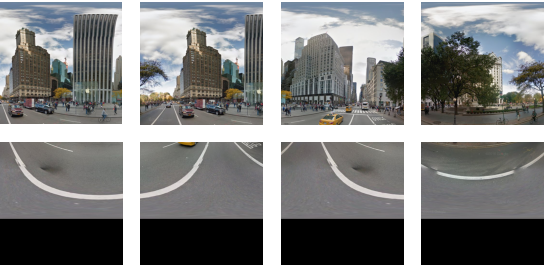}
      \caption{Small images with level 2 quality}
      \label{fig:pano1}
      \end{subfigure} %
      ~
      \begin{subfigure}[t]{0.48\linewidth}
\centering
      \includegraphics[width=\linewidth]{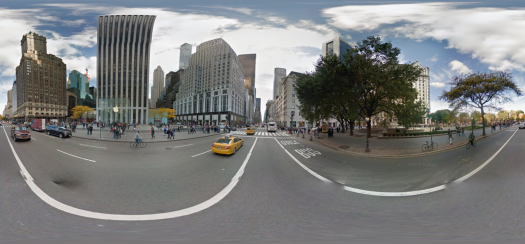}
      \caption{Panoramic photo stitched from the small images}
      \label{fig:pano2}
      \end{subfigure} %
\caption{Google street panoramic view}
\end{figure}

Fig.~\ref{fig:pano1} and Fig.~\ref{fig:pano2}
illustrate an example of small images and its corresponding panoramic photo. Fig.~\ref{fig:pano1} contains 8 512$\times$512 (pixels) figures (level 2).
These figures form a matrix where the adjacent images have some overlaps.
It implies that they can be further divided into two groups of four
images which can be stitched into two larger photos and they can be used as the base images when constructing Fig.~\ref{fig:pano2}.
If multiple adjacent images  are handled by one service node, this node can stitch these small images into an intermediate one
and reduce the computation at the end node.

\begin{table}[ht]
\caption{Google Street View Workloads}
 \centering
 \small
 \begin{tabular}{c l}
  \hline\hline
  Number & Location \\
  \hline
  1    &    Apple Store Fifth Avenue, NYC, NY \\
  \hline
  2    &    Metropolitan Museum of Art(indoor), NY \\
  \hline
  3    &    San Francisco Fishermans Wharf, CA\\
  \hline
  4    &    Fremont Sunday Flea Market, Seattle, WA\\
  \hline
  5    &    Capitol Hill, Washington, D.C.\\
  \hline
  6    &    Miami Beach, Miami, FL\\
  \hline
  7    &    Sydney Opera House, Sydney, Australia\\
  \hline
  8    &    Taj Mahal, Burhanpur, India\\
  \hline
  9    &    Palace of Versailles, Versailles, France\\
  \hline
  10    &    The Colosseum, Rome, Italy \\
  \hline
  \hline
\label{table:locations}
\end{tabular}
\end{table}

\vspace{-0.3in}
\begin{figure}[ht]
\centering
\includegraphics[width=2.5in]{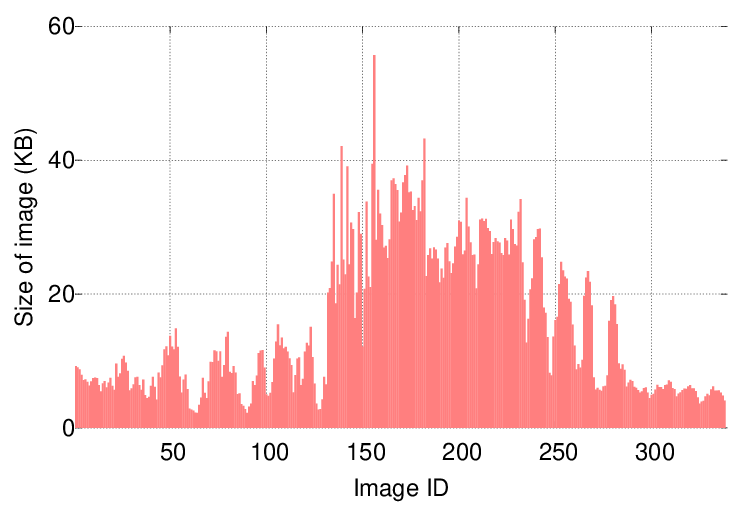}
\caption{Size of each small images in location 4 level 5}
\label{fig:small-images}
\end{figure}

\begin{figure*}[h]
   \centering
      \begin{subfigure}[t]{0.48\linewidth}
\centering
      \includegraphics[width=\linewidth]{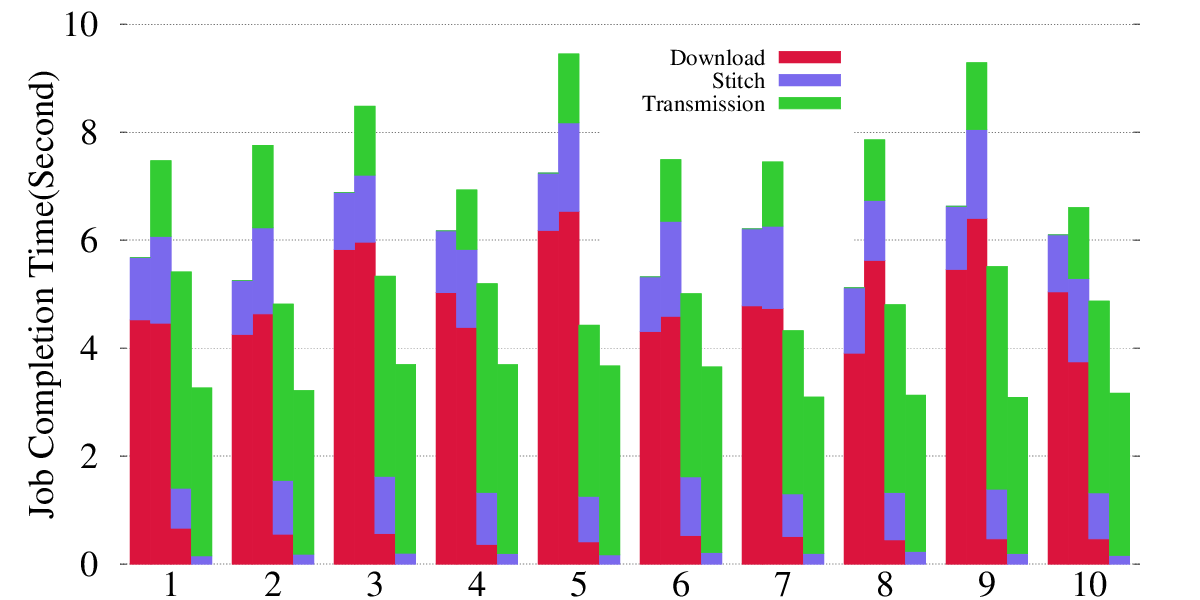}
      \caption{Comparison with one user}
      \label{fig:solo}
      \end{subfigure} %
      ~
      \begin{subfigure}[t]{0.48\linewidth}
\centering
      \includegraphics[width=\linewidth]{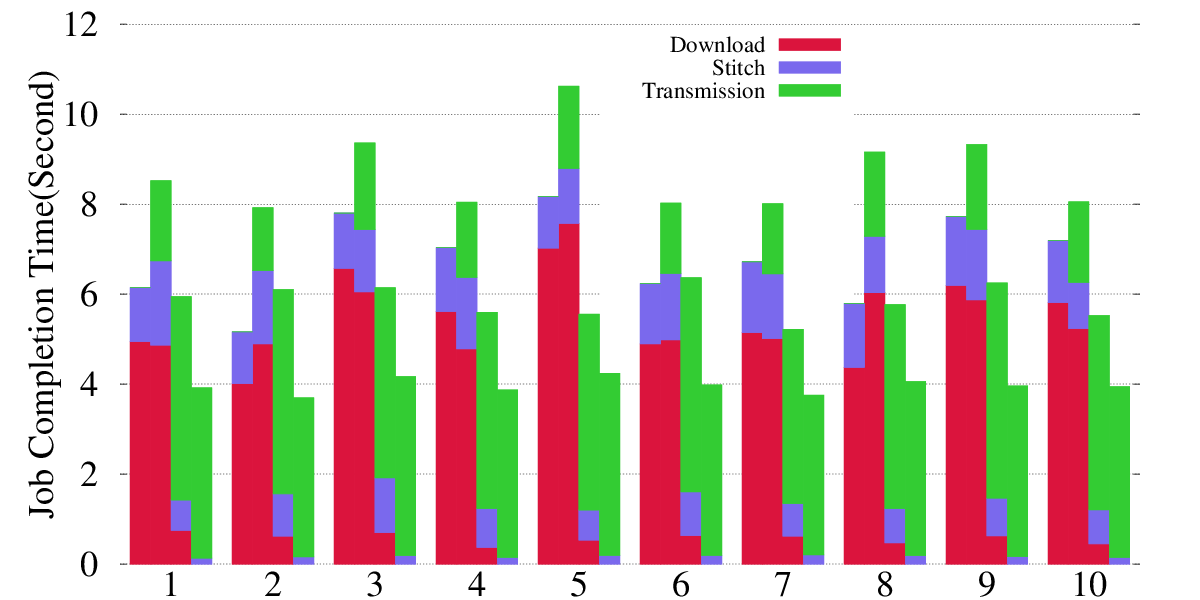}
      \caption{Comparison with two users}
      \label{fig:multiple}
      \end{subfigure} %
\caption{Average job completion time at different locations with level 4 resolution (91 small images)}
\end{figure*}

Table~\ref{table:locations} shows the 10 different locations that we used as the workloads for our \sol~system.
At each of the locations, we ran the experiments with 10 steps to simulate the moving forward action.
We tested all 5 levels at each step. Since whoever uses street view needs to download small images first, the size of each small image is an important metric. Fig.~\ref{fig:small-images} presents the sizes of tasks at location 4, level 5. The pictures with neighboring IDs are adjacent to each other. As we can see from the figure, the adjacent images have similar sizes because
the cameras that took these photos are geographically near each other with slightly different angles. Allocating adjacent images as a group to a node provides
benefits to the system. The reason lies in the fact that similar sizes result in a good alignment on service node and these images can be stitched into larger
one.


\subsection{System Implementation}
We implement \sol~system on commercial mobile devices and public clouds to build our testbed.
Introducing the heterogeneity into the testbed, it consists
of 3 mobile phones(iPhone 6, Google Nexus 5 and Huawei Mate 9),
3 pads (iPad Air, Samsung Galaxy Tab S2 and Google Nexus 7), and a Raspberry PI (runs Ubuntu) as the users and 3 Cloudlab~\cite{cloudlab} virtual machines as cloud service providers.
In addition, some of the users can connect to a Linksys WRT1900AC router with OpenFlow. In the system, \oashort~runs on all the users and \eashort~operates on the router.
All the participating nodes can specify several parameters.

\subsection{Performance Evaluation}
In this subsection, we present the results from both experiments on the testbed and simulations.

\subsubsection{Experiment results}
Recall that the main objective of \sol~is to complete the job with minimized time overhead and a given budget. The budget for a particular
job is given. However, the budget for each split task is not. In our experiments, we assign a divided budget to a task according to its size. Assuming the budget is
$Total_B$ there are
$n$ split tasks, for $i^{th}$ task,
its budget is $size_i/\sum_{i=1}^n size_i \times Total_B$.

To better evaluate \sol, we compare it with three different settings.
{\bf oSelf}: the user will complete the job itself, no offloading.
{\bf oNearby}: offloading all the tasks to a nearby service node which can be reached through Ad-hoc network or \eashort.
{\bf oCloud}: offloading all the tasks to remote cloud servers through \eashort.

\begin{figure*}[h]
   \centering
      \begin{minipage}[t]{0.50\linewidth}
\centering
      \includegraphics[width=\linewidth]{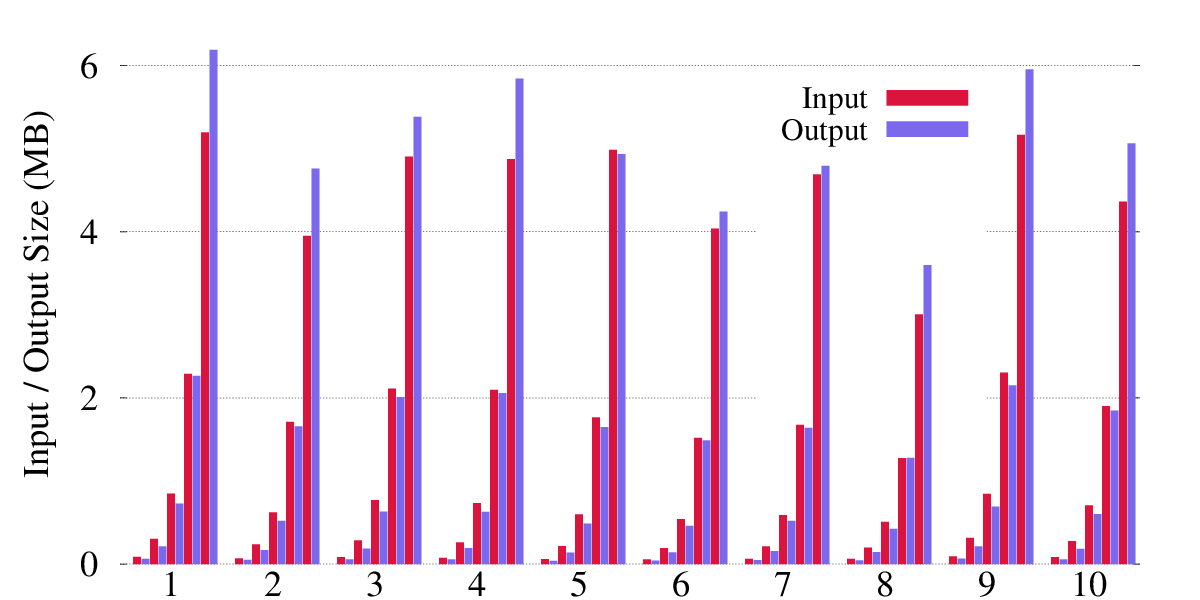}
      \caption{Input and output size on level 1-5 at location 1-10}
      \label{fig:size-location}
      \end{minipage} %
      ~
      \begin{minipage}[t]{0.46\linewidth}
\centering
      \includegraphics[width=\linewidth]{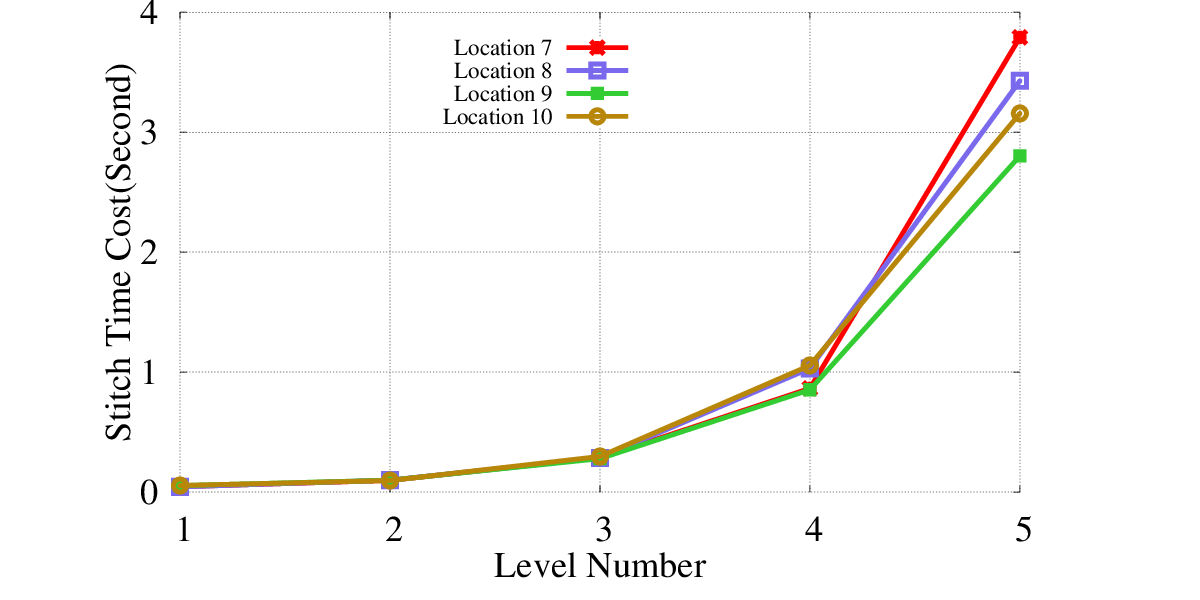}
      \caption{Stitching time cost at each level}
      \label{fig:stitch}
      \end{minipage} %
\end{figure*}

In the experiments, we use Bluetooth or WiFi Direct to construct an Ad-hoc network.
Fig.~\ref{fig:solo} and Fig.~\ref{fig:multiple} plot the results for the single user and two users settings with level 4 resolution.
For each of the locations, we run the experiments at 10 steps and calculate the average completion time.
At each location, there are four columns that represent oSelf, oNearby, oCloud, and \sol, respectively. From the figures, we have several findings.
Firstly, we can see that the completion time
of oSelf does not contain transmission because the user downloads all the raw data (small images) itself and does not request an offloading.
On the other hand, the completion time of \sol~does not includes download which is due to using \sol, it does not need raw data; instead, the nearby users and/or remote
servers will send the processed data to it during transmission time. Secondly, in both settings, \sol~achieves the shortest completion time.
For example, with a single user, \sol~completes the job 3.085s, while, oSelf, oCloud and oNearby consume 6.628s, 9.287s, 5.514s, respectively.
The reason is that \sol~introduces multiple service providers including nearby users and clouds. In \sol, the job has been split into multiple tasks which
be processed in parallel on different nodes. The parallel processing accelerates transmission since WiFi Direct has a much higher rate than regular WiFi.
Finally, the downloading time contributes to the majority of total completion time. In the figures, the downloading cost is not stable in a wireless setting. The duration of downloading starts from the first image until
the last one. It requires all small images to be downloaded to construct a panoramic view.
If any one of them were delayed it would result in a late start on the stitching process.
The user can download multiple images simultaneously. However, the larger number of concurrent tasks, the more likely to get one of them delayed.

The number of split tasks, the size of input and output is another factor that has an impact on the total cost.
Fig.~\ref{fig:size-location} illustrates the input and output sizes. At each location, the five clustered columns represent level 1-5.
From level 1-4, the input size is larger than the output. For example, at location 3, the input and output sizes for level 1 are, 85.053KB and 57.300KB, which reduced 32\%;
the reduction of level 2, 3 and 4 is 28\%, 17\%, and 5\%. These reductions come from the overlaps between the small images. When stitching, the overlaps will be removed.
The reduction is lower along with the increase of resolution because the algorithm not only removes but also introduces some metadata on each image,
such as orientation. The metadata dominates the change of sizes along with the number of small images. From level 4 to 5, this number increases from 91 to 338, and
the resulting output does not decrease in size but increases 14\%.


As the final step, stitching is another factor that contributes to the total cost. Stitching is a computationally intensive job and relys on the computation of CPU.
In our experiments with the same number of images, the server has the fastest stitching time. For example, at location 5 level 4, the stitching times for the server,
iPhone 6, and Nexus 7 are 671ms, 1152ms, and 1592ms.

Besides CPU, which is a feature specific to each user,
the number of images is the main factor under control by the system.
In Fig.~\ref{fig:solo} and Fig.~\ref{fig:multiple}, the stitching time of \sol~has been significantly reduced. For example, in a two user setting at location 5, \sol~costs 149ms for stitching and
others use 1421ms, 1592ms,864ms, respectively. The reason is that \sol~does not need to stitch all 91 small images in level 4. Depending on the selected nodes
for offloading on the user side, it only needs to stitch a limited number of images, e.g. 2-4 in our experiments. Fig.~\ref{fig:stitch} shows the stitching time
cost at each level. It is a clear trend that the cost increases along with the number of images.

\subsubsection{Simulation results}
To evaluate on a large scale network, we conduct simulations to test the performance of \sol. Our goal is to study the impact of the number of users on the system performance, concerning completion time.
In our simulations, we distinguish different service providers by several parameters discussed in~\ref{sec:problem}.
We set the value of parameters based on the intensive experiments above. Recall that a user can reach three types
of service nodes which are: (Type 1) cloud servers registered \eashort, (Type 2) devices connected to \eashort, (Type 3) nearby users discovered through Ad-hoc networks.
The table~\ref{table2} shows the values we derived from experiments. The parameters are randomly selected within the intervals.
Note that in the simulation we consider the static case where $Q_{j,n}=1, \forall j,n$ and communicational costs to be the same towards all nodes, i.e., $C_j^t$ is the same for all nodes $ v_j $.
The environment consists of 10 cloud service providers and 30 mobile devices, 15 of them can be reached through
\eashort~and the other 15 can be accessed through Ad-hoc networks. We use the same workloads and job splitting scheme as in the experiments.

\begin{table}[htb]
    \centering
    \caption{Parameters derived from experiments}

        \begin{tabular}{ | c | c | c | c | }
            \hline
            Parameters      & Type 1    &  Type 2     &     Type 3 \\ \hline
            $D_j^c$         & (0,0.05]    &  [0.01,0.1]    &    [0.01,0.1] \\ \hline
            $D_j^t$         & [1,3]    &  [0.5,2]    &    [0.5,1] \\ \hline
            $C_j^c$         & [0,5]    &  [0,5]    &    [0,5]    \\ \hline
            $C_{j,B}$     & $[1,50]\times C_j^c$ & $[1,50]\times C_j^c$ & $[1,50]\times C_j^c$ \\ \hline
        \end{tabular}
    \label{table2}
    \vspace{-5pt}
\end{table}

\begin{figure*}[h]
   \centering
      \begin{minipage}[t]{0.48\linewidth}
\centering
      \includegraphics[width=\linewidth]{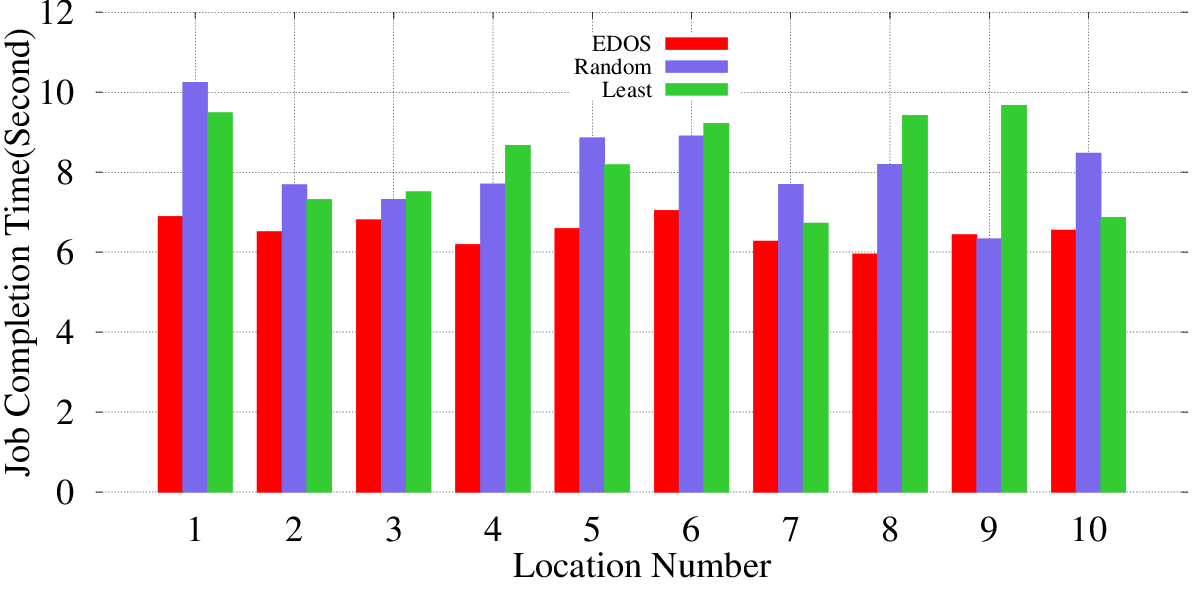}
      \caption{Job completion time with level 5 workloads}
      \label{fig:sim1}
      \end{minipage} %
      ~
      \begin{minipage}[t]{0.48\linewidth}
\centering
      \includegraphics[width=\linewidth]{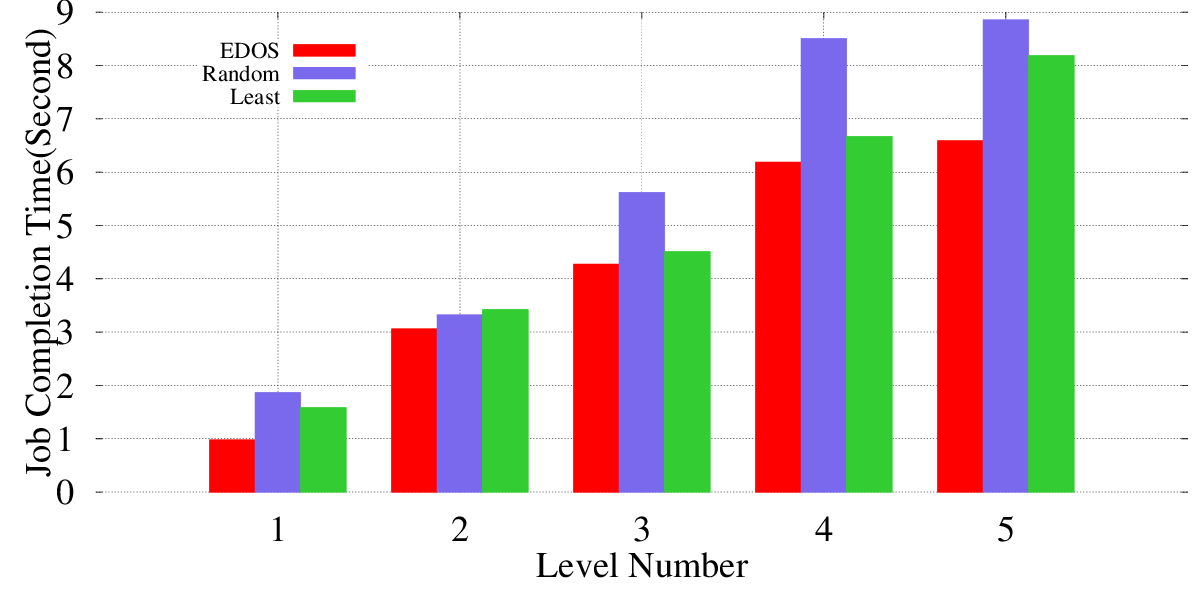}
      \caption{Job completion time at location 5}
      \label{fig:sim2}
      \end{minipage} %
\end{figure*}

In the simulation, we compare \sol~with two different task allocation schemes:
1) randomly selected service providers ($Random$);
2) always select available nodes with least cost ($Least$);
Fig.~\ref{fig:sim1} illustrates the simulation tests with level 5 that contains 338 split tasks.
As we can see from the figure, both the $Random$ and $Least$ solutions result in unstable completion time with similar input and output sizes. For the $Random$ approach, it is caused by the fact that the user does not have the control of which service nodes to offload.
If one node drains the total budget, the rest of the tasks have to be executed locally (no cost for the user).
On the other hand, the $Least$ solution tries to minimize the cost for tasks. However, if there exists a service provider that offers low cost, but extremely large delay, as shown at location 8, the job completion time would be much larger.

Next, we study the impact of the number of tasks in the system. Fig.~\ref{fig:sim2} plots the job completion time at location 5 with level 1-5.
Recall that, at each level, the number of tasks is 2, 8, 28, 91, and 338. As shown in the figure, \sol~outperforms
the other two solutions substantially at level 4 and 5.
The performance gain of \sol~is smaller at level 1 to 3, because the number of tasks is limited and it is more likely that 1-2 service providers hold all the tasks.

\vspace{-0.07in}
\section{Conclusion}
This paper develops \sol, a cost-aware hybrid offloading system with edge assistance.
\sol~is based on the \eashort~that runs on the routers/towers and \oashort,~which operates
on the users' devices. We present service discovery protocols based on both
infrastructure-based networks and Ad-hoc networks.
The user splits a job into multiple tasks and allocates them to appropriate service providers according to user-specified
constraints and to reduce the job completion time.
We prototype \sol~on commercial mobile devices and evaluate it with both experiments on a small-scale testbed and simulations for a large-scale setting.
The results show that \sol~system is effective and efficient for
offloading jobs.

\noindent{\bf Acknowledgement:} This project was supported by National Science Foundation grant CNS-1527336 and TCNJ SOS Mini grant.

\bibliographystyle{unsrt}
\bibliography{routing}
\end{document}